\documentclass[]{spie}  

\usepackage{epsfig}
\usepackage{graphicx}
\usepackage{amssymb}
\usepackage{nicefrac}
\usepackage{cool}
\usepackage{mathrsfs}
\usepackage{units}
\usepackage{amsmath}
\usepackage{footmisc}
\usepackage{hyperref}
\usepackage{hyphenat}
\Style{DSymb={\mathrm d},IntegrateDifferentialDSymb=\text{d}}

\def\piper{\textsc{Piper}}
\def\bicep2{\textsc{Bicep2}}

\def\abs{\textsc{Abs}}
\def\scuba2{\textsc{Scuba2}}

\def\arcade1{\textsc{Arcade1}}
\def\planck{\textsc{Planck}}

\title{The Primordial Inflation Polarization Explorer (PIPER)}

\author{Justin Lazear\supit{a},
        Peter~A.~R.~Ade\supit{b},
        Dominic~Benford\supit{c},
        Charles~L.~Bennett\supit{a},
        David~T.~Chuss\supit{c},
        Jessie~L.~Dotson\supit{d},
        Joseph~R.~Eimer\supit{a},
        Dale~J.~Fixsen\supit{c},
        Mark~Halpern\supit{e},
        Gene~Hilton\supit{f}
        James~Hinderks\supit{c,i},
        Gary~F.~Hinshaw\supit{e},
        Kent~Irwin\supit{g},
        Christine~Jhabvala\supit{c},
        Bradley~Johnson\supit{h},
        Alan~Kogut\supit{c},
        Luke~Lowe\supit{c,k},
        Jeff~J.~McMahon\supit{i},
        Timothy~M.~Miller\supit{c},
        Paul~Mirel\supit{c,k},
        S.~Harvey~Moseley\supit{c},
        Samelys~Rodriguez\supit{c,j},
        Elmer~Sharp\supit{c},
        Johannes~G.~Staguhn\supit{a,c},
        Eric~R.~Switzer\supit{c},
        Carole~E.~Tucker\supit{b},
        Amy~Weston\supit{c,j}, and
        Edward~J.~Wollack\supit{c}
\skiplinehalf
\supit{a}Johns Hopkins University, Baltimore, MD, United States; \\
\supit{b}Cardiff University, Cardiff, Wales, United Kingdom; \\
\supit{c}Code 665, NASA Goddard Space Flight Center, Greenbelt, MD, United States; \\
\supit{d}NASA Ames Research Center, Moffett Field, CA, United States; \\
\supit{e}University of British Columbia, Vancouver, BC, Canada; \\
\supit{f}National Institute for Standards and Technology, Boulder, CO, United States; \\
\supit{g}Stanford University, Stanford, CA, United States; \\
\supit{h}Columbia University, New York, NY, United States; \\
\supit{i}University of Michigan, Ann Arbor, MI, United States; \\
\supit{j}{ADNET} Systems, Inc., Bethesda, MD, United States; \\  
\supit{k}Wyle {STE}, Houston, TX, United States  
}

\authorinfo{Send correspondence to Justin Lazear: jlazear@pha.jhu.edu\\
Copyright 2014 Society of Photo Optical Instrumentation Engineers. One print
or electronic copy may be made for personal use only. Systematic reproduction
and distribution, duplication of any material in this paper for a fee or for
commercial purposes, or modification of the content of the paper are
prohibited.}

\begin{document}
\maketitle

\begin{abstract}
The Primordial Inflation Polarization Explorer (\piper) is a balloon-borne cosmic microwave background (CMB) polarimeter designed to search for evidence of inflation by measuring the large-angular scale CMB polarization signal.
\bicep2\ recently reported a detection of B-mode power corresponding to the tensor-to-scalar ratio $r = 0.2$ on $\sim\!\!2$ degree scales. If the \bicep2\ signal is caused by inflationary gravitational waves~(IGWs), then there should be a corresponding increase in B-mode power on angular scales larger than 18 degrees.
\piper\ is currently the only suborbital instrument capable of fully testing and extending the \bicep2\ results by measuring the B-mode power spectrum on angular scales $\theta = \sim\!\!0.6^\circ$ to $90^\circ$, covering both the reionization bump and recombination peak, with sensitivity to measure the tensor-to-scalar ratio down to $r = 0.007$, and four frequency bands to distinguish foregrounds.
\piper\ will accomplish this by mapping 85\% of the sky in four frequency bands (200, 270, 350, 600 GHz) over a series of 8 conventional balloon flights from the northern and southern hemispheres. The instrument has background-limited sensitivity provided by fully cryogenic (1.5~K) optics focusing the sky signal onto four $32\!\!\times\!\!40$-pixel arrays of time-domain multiplexed \nohyphens{Transition-Edge~Sensor~(TES)} bolometers held at 140~mK. Polarization sensitivity and systematic control are provided by front-end Variable-delay Polarization Modulators (VPMs), which rapidly modulate only the polarized sky signal at 3~Hz and allow \piper\ to instantaneously measure the full Stokes vector ($I, Q, U, V$) for each pointing. We describe the \piper\ instrument and progress towards its first flight.
\end{abstract}



\section{INTRODUCTION}
\label{sec:intro}

\begin{figure}
   \begin{center}
       \includegraphics[width=\textwidth]{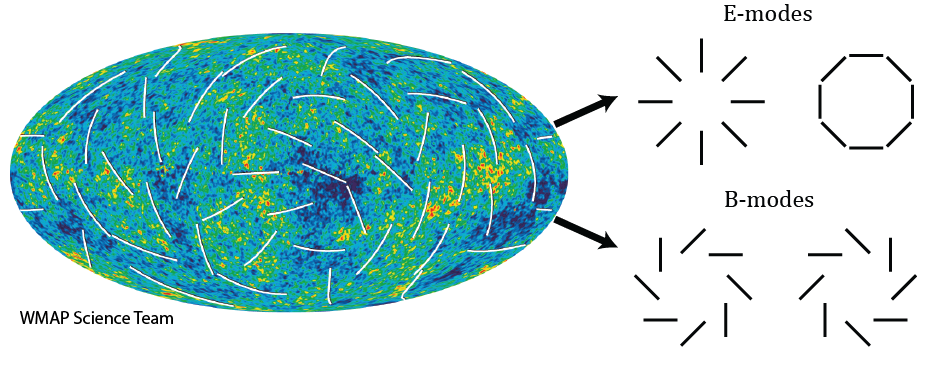}
   \end{center}
   \caption[E- and B-modes]{The CMB anisotropy polarization map may be decomposed into curl-free even-parity E-modes and divergence-free odd-parity B-modes. Primordial B-modes are only created by tensor perturbations (inflationary gravitational waves).}
   \label{fig:ebmodes}
\end{figure}

The current leading theory in cosmology predicts that shortly after the Big Bang, the universe underwent an epoch of rapid expansion, called inflation. Inflation would have created a background of gravitational waves, which in turn would have imprinted themselves on the cosmic microwave background (CMB) as a characteristic spatial pattern in the CMB polarization\cite{hu_cmb_1997,kamionkowski_statistics_1997}. The spatial patterns of the CMB polarization can be decomposed into E- and B-modes (Fig.~\ref{fig:ebmodes}), named in analogy to the curl- and divergence-free nature of the electric and magnetic fields, respectively. Thomson scattering of radiation with a quadrupolar distribution is the mechanism by which polarization may be imprinted onto the CMB.

\begin{figure}
   \begin{center}
       \includegraphics[height=7cm]{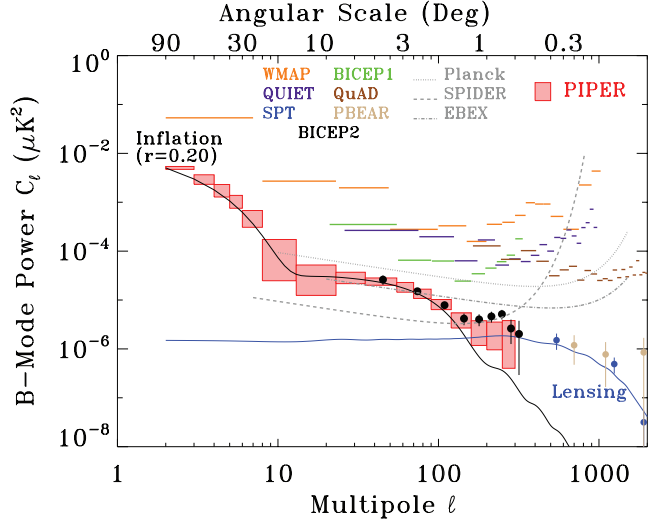}
   \end{center}
   \caption[BB spectrum]{The B-mode power spectrum assuming a tensor-to-scalar ratio $r = 0.2$ from inflationary gravitational waves and gravitational lensing. The quantity being measured $C_\ell^\mathrm{BB}$ is plotted, rather than $\ell (\ell + 1)C_\ell^\mathrm{BB}/2\pi$. Theories of inflation generically predict a rise in power at $\ell < 10$. \piper\ will measure the shape of the spectrum for $2 \leq \ell < 300$, encompassing the ``reionization bump'' at low $\ell$, the ``recombination bump'' at $\ell \sim 80$, as well as the lensing signal at $\ell > 200$.}
   \label{fig:bb}
\end{figure}

A quadrupolar distribution of incident radiation at recombination could be created by quadrupolar temperature anisotropy or differential Doppler shifting by gravitational waves. Anisotropy due to statistical temperature fluctuations are scalar perturbations and can create only E-mode patterns. Gravitational waves are tensor perturbations that create both E- and B-mode patterns. If the only source of gravitational waves prior to recombination was inflation, then inflationary gravitational waves (IGWs) would be the only expected source of B-modes from the early universe. IGWs in the recombination epoch would have produced a characteristic peak in the B-mode power spectrum near angular scale $\ell \sim 180^\circ/\theta \sim 80$ (Fig.~\ref{fig:bb}) with a blackbody spectral distribution.

B-mode patterns may also have been created after recombination by Thomson scattering during the epoch of reionization, gravitational lensing of CMB photons by intervening matter, synchrotron emission of cosmic ray electrons, and polarized interstellar dust. Thomson scattering during reionization produces a similar effect to the recombination peak, but at larger angular scale ($\ell < 10$) since the horizon was closer by the time of reionization. Gravitational lensing converts some E-mode power to B-mode power without affecting the spectral distribution, and dominates the IGW signal for $\ell \gtrsim 200$. Neither synchrotron emission nor polarized dust emission have blackbody spectral distributions, and must be cleaned from the maps; they are not included in Fig.~\ref{fig:bb}.

B-mode power is characterized by the ratio of the tensor power contributions to the quadrupole with the scalar power contributions, the tensor-to-scalar ratio $r$. The tensor-to-scalar ratio is related to the amplitude of the gravitational wave background, and thus the energy scale of inflation. A detection of B-modes with the characteristic shape shown in Fig.~\ref{fig:bb} is definitive evidence for inflation. For a given $r$, inflation predicts and sets the scale of both a recombination peak at $\ell \sim 80$ and a reionization bump at $\ell < 10$. \bicep2{} recently reported\cite{collaboration_bicep2_2014} measurements of B-modes at the $r = 0.20^{+0.07}_{-0.05}$ level based on a power spectrum between $30 < \ell < 150$ (the recombination peak) in a 380 sq.\ deg.\ low-foreground patch of sky with a single 150 GHz frequency band. This is in tension with the constraint of $r < 0.11$ set by \planck\cite{collaboration_planck_2013} temperature maps combined with other measurements, using the simplest models of inflation. Further measurements of the B-mode spectrum will be important for reconciling this tension and for assessing the foregrounds.

A critical measurement is mapping the B-mode spectrum between $2 \leq \ell < 300$, constraining both the reionization bump and recombination peak. It is also necessary to map the sky in multiple spectral frequency bands to allow the maps to be cleaned of foregrounds. Such a measurement requires excellent systematic control and high sensitivity to integrate down to the required $\unit[\sim\!\! 10]{nK}$ scale.

A number of suborbital missions are targeting the B-mode spectrum at the recombination peak\cite{filippini_spider:_2010,reichborn-kjennerud_ebex:_2010}, but \piper\ is the only suborbital instrument in the near future that will be able to map the full B-mode spectrum for $\ell < 300$, including the reionization bump, in multiple frequency bands. Flights from both hemispheres will enable mapping of 85\% of the sky, allowing \piper\ to reach $\ell$ as low as 2. Four frequency bands (200, 270, 350, 600 GHz with 30\%, 30\%, 16\%, and 10\% bandwidths) will provide the necessary frequency coverage to treat foregrounds. Front-end \nohyphens{Variable-delay~Polarization~Modulators~(VPMs)}\cite{chuss_variable-delay_2014} will rapidly modulate only the polarized sky signal, allowing \piper\ to reject systematics. Fully cryogenic optics and 140~mK \nohyphens{transition-edge sensor~(TES)} bolometers provide background-limited performance, giving \piper\ instantaneous sensitivity better than $\unit[2]{\mu K\ s^{1/2}}$. In a series of 8 conventional overnight balloon flights alternating between northern and southern hemispherse, \piper\ will be able to map 85\% of the sky in 4 frequency bands, covering $2 \leq \ell < 300$ with sufficient sensitivity to reach $r < 0.007$, facilitating a $30\sigma$ test of the \bicep2\ results.

\begin{figure}
   \begin{center}
       \includegraphics[height=10cm]{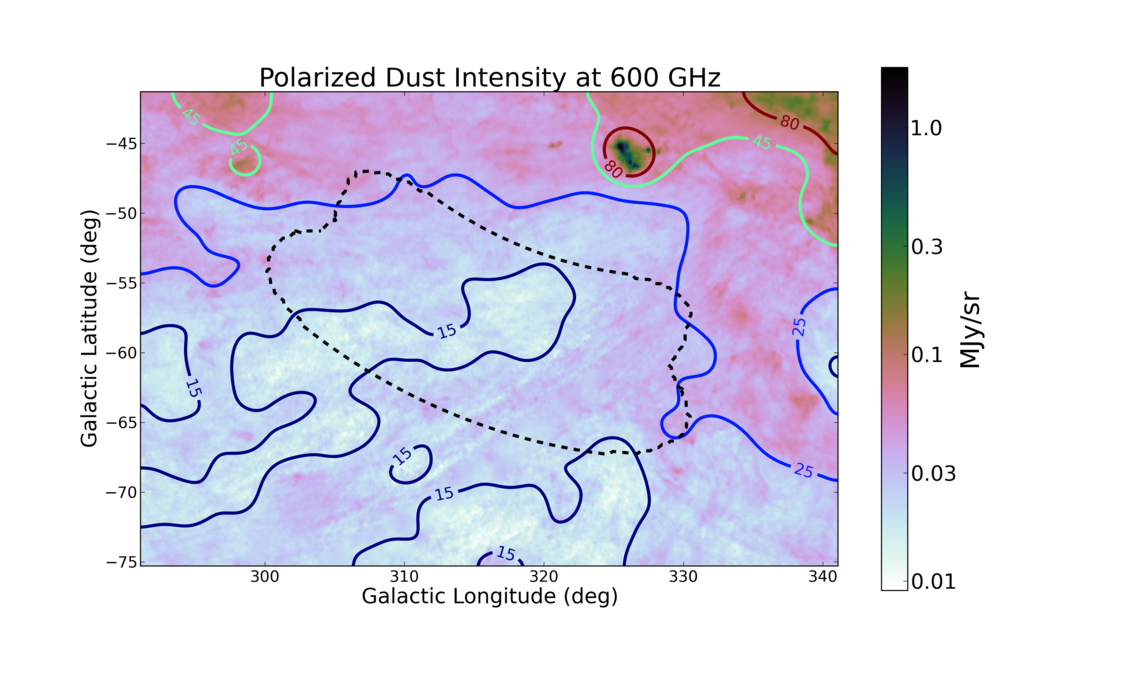}
   \end{center}
   \caption[PIPER Dust Map]{Projected polarized dust intensity at 600 GHz in the \bicep2\ region (dashed line) assuming a 10\% polarization fraction\cite{finkbeiner_extrapolation_1999}. S/N contours are shown for $\ell \sim 80$ after a single 600 GHz flight. \piper\ will be able to measure the polarized dust foreground to a S/N of better than 10 even for regions of low dust intensity and even for polarization fractions as low at 10\%.}
   \label{fig:dust}
\end{figure}

Recent studies have suggested that the polarized dust foregrounds may be a significant contributor to the \bicep2\ B-mode detection\cite{mortonson_joint_2014,flauger_toward_2014}. Though current \planck\ polarized dust maps do not cover the \bicep2\ region, the maps suggest that low dust intensity regions of the sky may be more strongly polarized than previously expected\cite{collaboration_planck_2014}. Thus, even though \bicep2\ observed in a region of low total dust intensity, it may still have a significant polarized dust foreground. \piper\ will be able to unequivocally resolve the polarized dust foreground, even in low intensity regions such as the \bicep2 region (Fig.~\ref{fig:dust}).

\section{INSTRUMENT OVERVIEW}


\piper\ is a balloon-borne millimeter and sub-millimeter polarimeter designed to measure the B-mode signature from IGWs in the CMB. It utilizes a twin-telescope design to simultaneously measure all components of the full Stokes vector ($I, Q, U, V$) at each pointing. A front-end VPM on each telescope modulates the sky signal at 3 Hz between the local Stokes U and V, allowing for effective systematic control. Fully-cryogenic optics minimize emission from optical elements and, together with TES bolometers held at 140 mK, makes photon statistics uncertainty from the sky signal itself the dominant noise source. At this noise floor, the only way to gain more sensitivity is to collect more photons. To this end, \piper\ employs four $32\!\!\times\!\!40$ pixel arrays, providing $<\!\unit[2]{\mu K\ s^{1/2}}$ instantaneous polarization sensitivity with a 21 arcmin beam.

\piper\ will fly 8 times: 4 times from Ft. Sumner, NM and 4 times from Alice Springs, Australia. \piper's design does not require any Antarctic flights, only conventional flights. Each pair of flights covers 85\% of the sky in a single frequency band, permitting measurement down to $\ell = 2$. Frequency bands are determined by a bandpass filter\cite{ade_review_2006} immediately in front of the detectors; anti-reflective (AR) coatings and the throw of the VPM optimize the efficiency of the system. The detectors are broadband and will be re-used with each flight. This provides unparalleled flexibility for the \piper\ instrument: it may adjust frequency bands of subsequent flights based on the most up-to-date data to maximize science returns, allowing it to optimize between foreground rejection and science signal depth.

\section{OPTICS}
\label{sec:optics}

\begin{figure}
   \begin{center}
       \includegraphics[width=\textwidth]{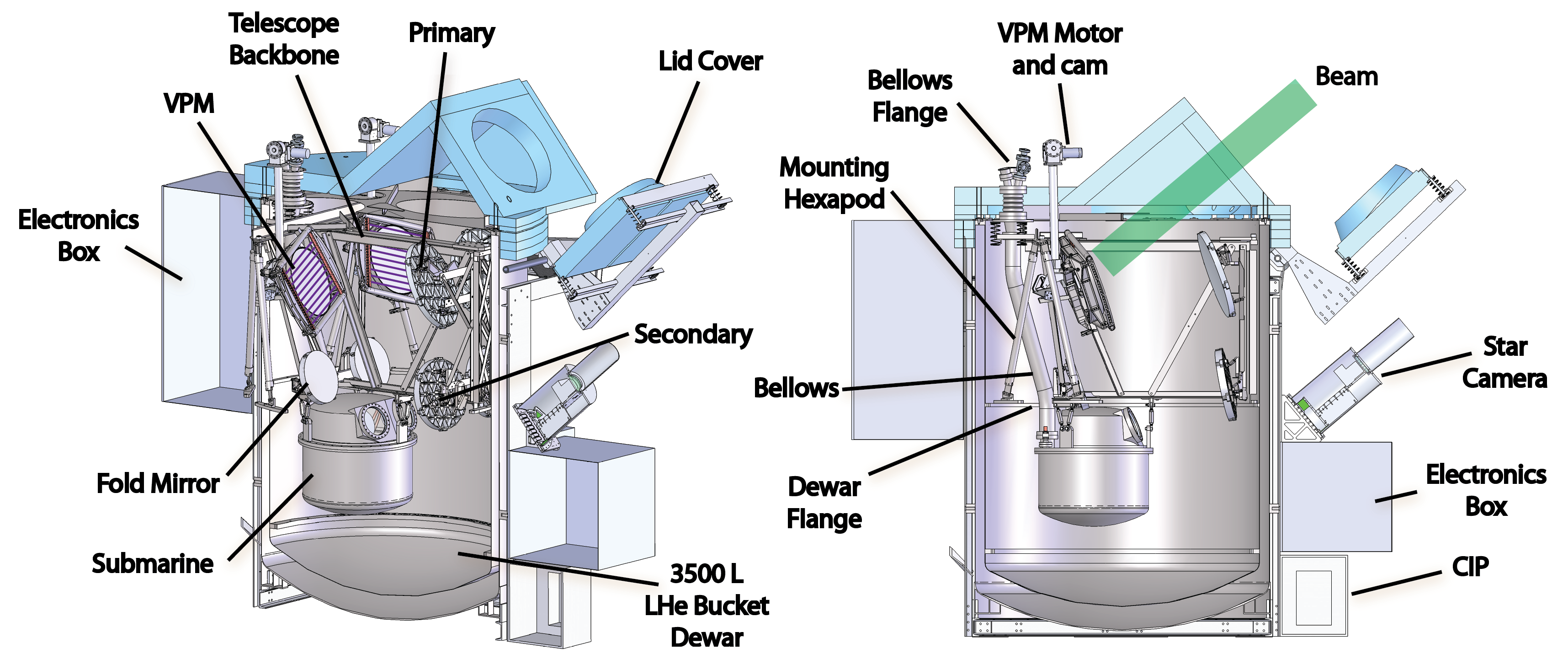}
   \end{center}
   \caption[Optical layout]{The twin telescopes are mounted by a fully stainless steel frame in an open-aperture 3500 liter LHe bucket dewar. The telescope mounts and is registered to a rigid backbone; the entire backbone is mounted into the dewar. Only elements of one telescope are labeled; the two telescopes are sagittal mirrors of one another. The first optical element the sky sees is the VPM, ensuring instrument generated polarization cannot be modulated and become systematics. The VPM is actuated by a warm motor exterior to the dewar. A retractable lid cover protects the detectors on the ground and from the sun. A star camera provides absolute pointing information.}
   \label{fig:optics}
\end{figure}

The optical configuration\cite{eimer_primordial_2010} used in \piper\ is shown in Fig.~\ref{fig:optics}. The entire telescope is housed in a 3500 liter open-aperture liquid helium (LHe) bucket dewar that, after atmospheric pumping during the ascent to the 30 km float altitude, serves to keep all of the optical elements at $<\unit[\!1.5]{K}$ with the aid of superfluid LHe pumps. The photon noise from blackbody emission of optical elements held at 1.5 K is negligible, allowing \piper\ to operate very near the noise floor for a balloon-borne experiment. Compared to an instrument with a warm window, this is a reduction in noise power by a factor of 3 and a corresponding increase in mapping speed by a factor of 10. \piper\ leverages this increased speed to map, to the required sensitivity, 55\% of the sky in a single overnight conventional balloon flight.

A fully stainless steel telescope mount holds all of the optical elements. Because the mounting structure and optical bracketry are all made of the same material, they contract self-similarly under cooling so that the relative positions and shapes of the optics are preserved (up to a single scaling factor). The only exceptions are the aluminum mirrors due to the prohibitive cost of manufacturing stainless steel mirrors. The differential thermal contraction between the aluminum mirrors and stainless steel frame is well understood. All of the optical elements are mounted on and referenced to a single backbone. The entire system is inserted into the dewar as a unit, where the system is secured, without disturbing the optical alignment, by a hexapod supporting the backbone.

The optical system will be aligned on an external stand using a coordinate measuring machine (CMM) arm and reference geometry built into every optical element. Each reference point on each optical element must be located to within $\unit[\sim\!\!1]{mm}$; the CMM arm is capable of $\unit[40]{\mu m}$ absolute positioning.

\piper\ utilizes 6 monocrystalline high-resistivity silicon lenses: one for each telescope before the analyzer grid, and another for each beam path following the analyzer grids. To minimize losses from the lenses, they are AR-coated with a metamaterial stepped groove surface by the University of Michigan. This technique has been successfully used by ACTPol\cite{datta_large-aperture_2013}. The first coating provides a passband wide enough for both the 200 and 270 GHz bands; subsequent frequency bands will require new AR coatings.

\section{VARIABLE-DELAY POLARIZATION MODULATORS}
\label{sec:VPM}

\begin{figure}
   \begin{center}
       \includegraphics[width=\textwidth]{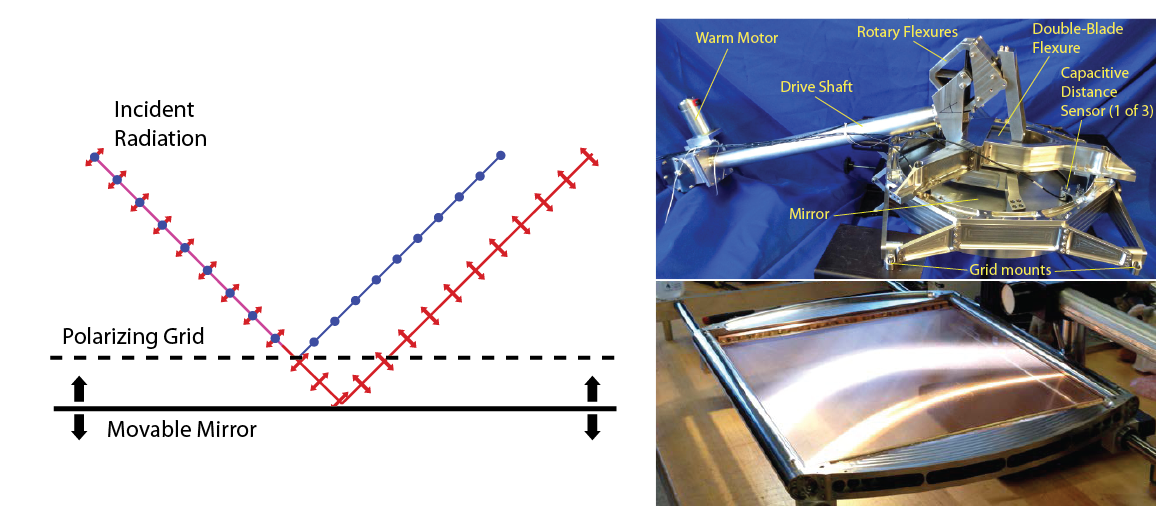}
   \end{center}
   \caption[VPM description]{A variable-delay polarization modulator (VPM) modulates only the Stokes $U$ and $V$ components of incident radiation.
   The component of polarization parallel to the wires is reflected at the grid and the component perpendicular to the wires is reflected at the mirror, inducing a phase delay between the two components and thereby rotating between $U$ and $V$. A fully stainless steel assembly actuates the mirror while constraining the wire grid. Three capacitive sensors measure the position of the mirror relative to the grid.}
   \label{fig:vpm}
\end{figure}

A Variable-delay Polarization Modulator (VPM)\cite{chuss_interferometric_2006,chuss_primordial_2010} consists of a wire grid in front of a movable flat mirror (Fig.~\ref{fig:vpm}). Polarized light incident on the grid is split into two components. The component polarized parallel to the wires in the grid is reflected at the grid; the component polarized perpendicular to the wire passes through the grid, reflects off the mirror, and recombines with the parallel component. The phase delay introduced between the two linear components rotates between the local Stokes $U$ and $V$, but leaves the $Q$ component unchanged. The output $U'$ is given by
\begin{equation*}
    U' = U\cos{\phi} + V\sin{\phi}
\end{equation*}
where $\phi$ depends linearly on the mirror-grid spacing. The polarization signal may be modulated at an arbitrary frequency with an arbitrary shape by appropriately modulating the mirror-grid spacing.

\piper\ uses a VPM as the first optical element for each of its twin telescopes. The VPMs are clocked $45^\circ$ relative to each other so that the local $U$ of one maps onto the sky $Q$ and the local $U$ of the other maps onto the sky $U$. In this way, \piper\ modulates the full polarization vector at the oscillation frequency of the VPM mirrors. The twin telescopes simultaneously measure all components of the total modulated polarization vector and the unmodulated intensity.

A design incorporating VPMs has a number of advantages. Since the VPMs are the first optical element, only the sky polarization signal is modulated while instrumental sources of polarized light are unmodulated. The sky signal may be isolated from instrumental noise by an appropriate demodulation, thereby greatly reducing contamination from the instrument. Additionally, there is minimal $Q-U$ mixing (resulting in $E-B$ leakage), since the VPMs modulate between local $U$ and $V$. The sky $V$ is expected to be negligible\cite{mainini_improved_2013}, so the modulated linear polarization signal will not be contaminated by circular polarization.

The 3 Hz oscillation of the VPM mirror, faster than the expected bandwidth of the sky signal, modulates the sky polarization signal at the same frequency. If the oscillation frequency is faster than the $1/f$ knee of the system then the signal band is unaffected by $1/f$ noise. The \abs\ instrument demonstrated the efficacy of rapid modulation\cite{kusaka_modulation_2014} using a rapidly rotating half-wave plate, noting a 30 dB reduced contribution of $1/f$ noise to the demodulated signal. \piper\ measures the mirror-grid separation and constructs the demodulation function from this data, removing the need for an extremely stable modulation frequency.

Simultaneous measurement of all components of the Stokes vector isolates the polarization sensitivity from the scan strategy and the beam shape. With the VPMs, each point on the sky needs to be observed only once in an arbitrary (but known) orientation, facilitating \piper's very simple scan strategy\cite{kogut_primordial_2012}. Without a reliance on crosslinking or boresight rotation, the effects of asymmetric beams are greatly mitigated, and \piper\ is immune to average signal level drifts between visits to a point in the sky.

For good performance, the VPMs rely on consistent parallel transport of the mirror relative to the grid, consistent wire spacing, and uniform flatness across the grid and mirror. The motion of the \piper\ VPMs results in mirror-grid relative tilting of less than 2.5 arcseconds across the full 1 mm throw. Each of the \piper\ VPM grids consist of a 39 cm clear aperture of $\unit[40]{\mu m}$ copper-plate tungsten wires separated by $\unit[117.0 \pm 5.7]{\mu m}$ that, when constrained by the grid flattener, are flat to $\unit[8.7]{\mu m}$\cite{chuss_variable-delay_2014}.

\section{DETECTORS}
\label{sec:detectors}


\piper\ will fly four arrays of $32\!\!\times\!\!40$-pixel backshort-under-grid (BUG) transition-edge sensor (TES) broadband bolometers\cite{jhabvala_kilopixel_2014} biased at 140 mK and coupled to a 100 mK bath. Each $\unit[1.135]{mm}\!\times\!\unit[1.135]{mm}$ pixel measures total incident in-band power; polarization sensitivity is provided by the upstream VPMs and analyzer grids. The reflective backshort is designed to maximize absorptivity in the science bands (200 and 270 GHz) while attentuating higher frequency bands where the foreground power is larger. Without this attenuation, the loading on each pixel would exceed the 1.2 pW saturation power in the higher frequency bands. The arrays are indium bump-bonded to a 2D superconducting quantum interference device (SQUID) time-domain multiplexer chip\cite{chervenak_superconducting_1999,irwin_-focal-plane_2004} and read out by Multi-channel Electronics (MCEs)\cite{battistelli_functional_2008}. The sky is imaged directly onto the detector arrays without the use of feedhorns in order to maximize throughput and pixel density.

The band definition is determined by bandpass filters at the front of the detector package. To maximize efficiency, the AR coatings on the lenses and the throw of the VPM mirror will be made to match the desired passband. An important feature of this design is that the frequency sensitivity is determined entirely by easily exchanged parts---the VPM throw, the AR coating, and the bandpass filter---and not by the far more expensive detectors. \piper\ will measure all 4 of its frequency bands using a single set of detectors, while changing out only the VPM cam, the lenses, and the bandpass filters to adjust the frequency band between flights.

In addition to reducing costs, this design provides flexibility. The first flights will cover the 200 and 270 GHz frequency bands, but the frequency bands of subsequent flights may be adjusted in the context of data from previous flights and other experiments. \piper\ will be able to easily optimize the frequency band choice for each flight in order to maximize science returns. The \piper\ detectors are capable of utilizing frequency bands as low as 150 GHz without a significant loss in NEP.

\begin{figure}
   \begin{center}
       \includegraphics[width=0.45\textwidth]{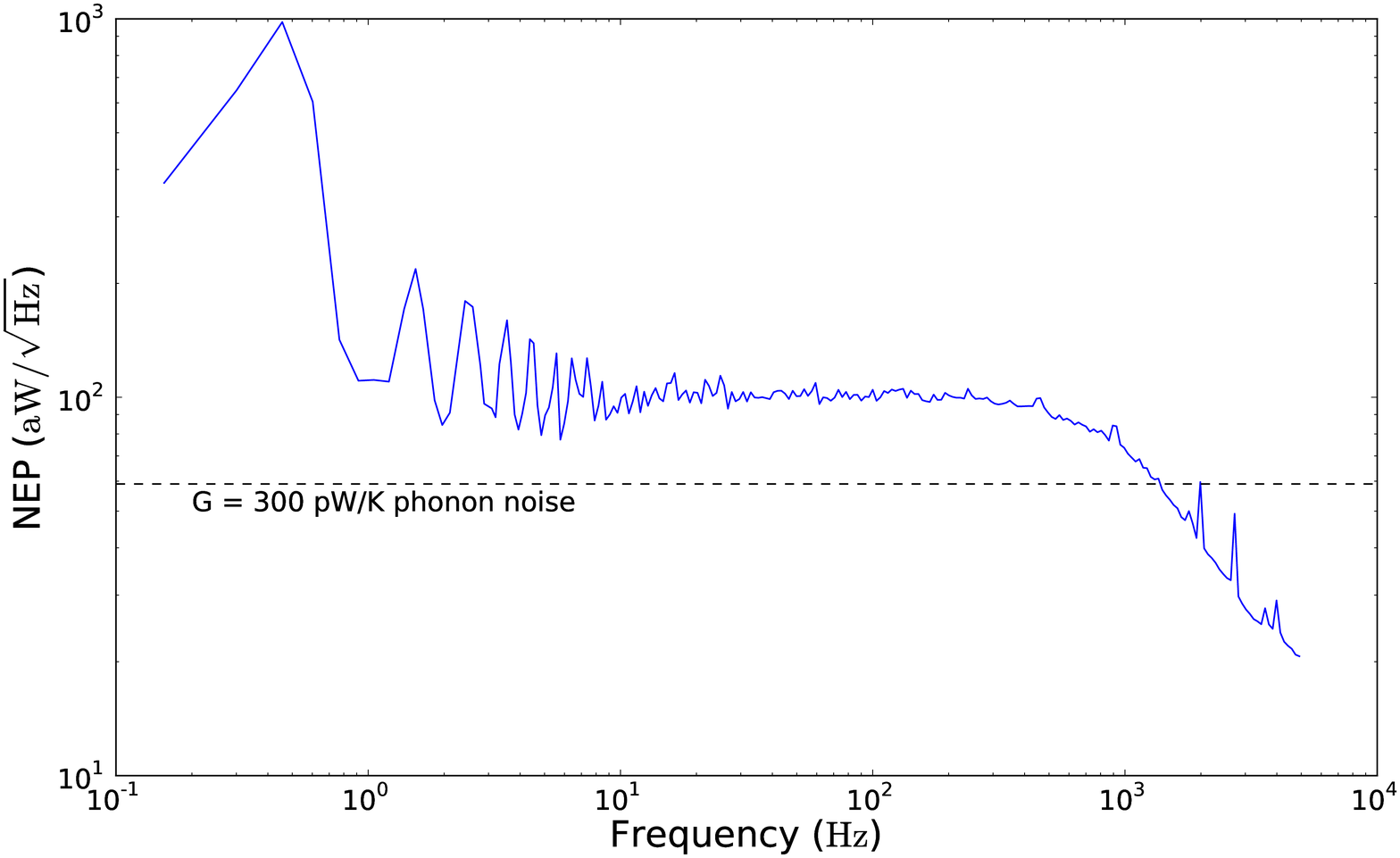}
       \includegraphics[width=0.45\textwidth]{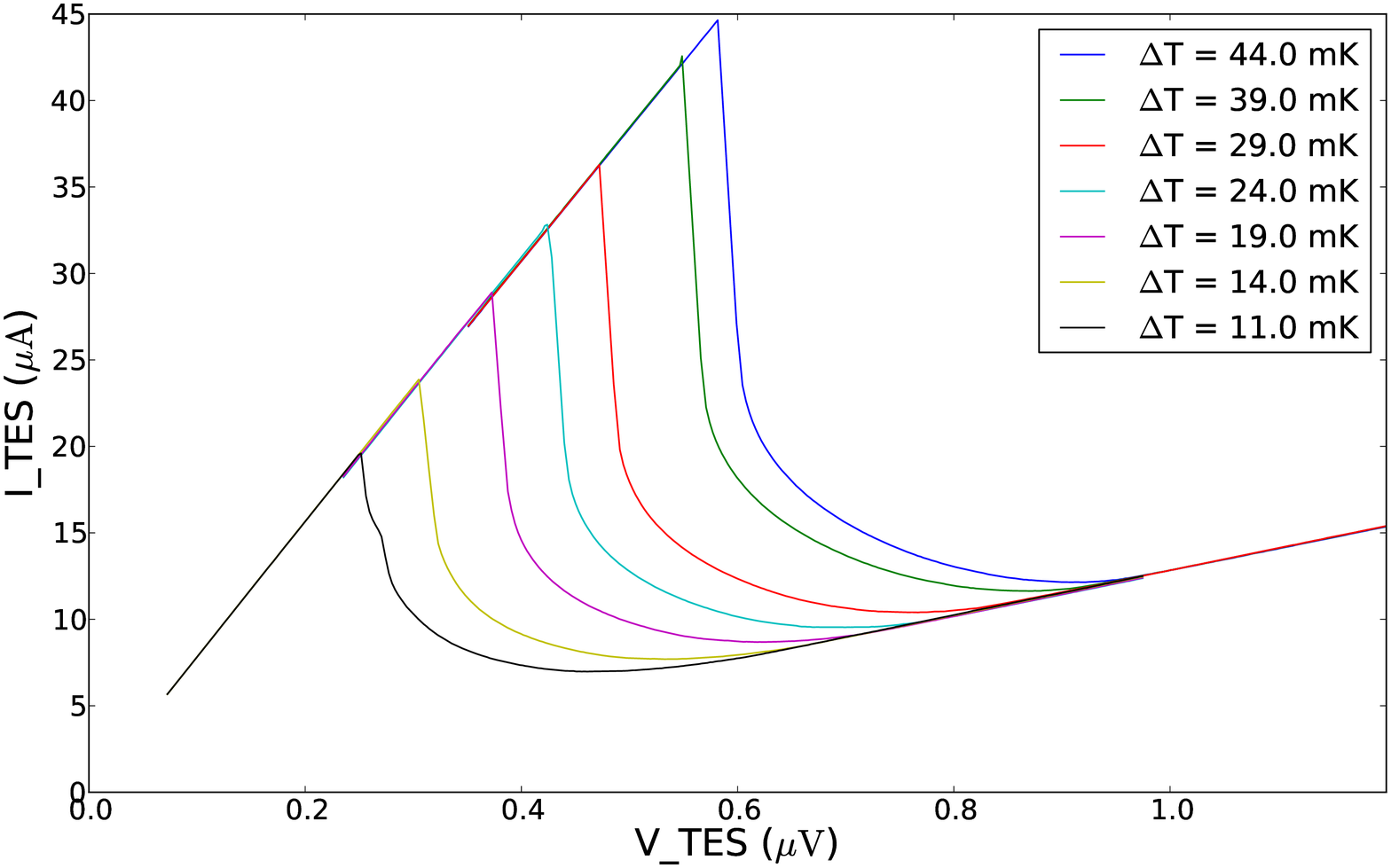}
   \end{center}
   \caption[SPTD]{(Left) An engineering detector with a larger $G$ and saturation power than the flight arrays shows NEP that is within a factor of 2 of the ideal phonon limit (dashed line). The noise is white in the modulation band and the roll-off is set by the electronics control loop bandwidth. (Right) A family of I-V curves for various $\Delta T = T_c - T_\mathrm{base}$ shows high quality transitions. The parasitic resistance is accounted for by the test circuit.}
   \label{fig:detectors}
\end{figure}

\section{CRYOGENICS}
\label{sec:cryogenics}

\begin{figure}
   \begin{center}
       \includegraphics[height=7cm]{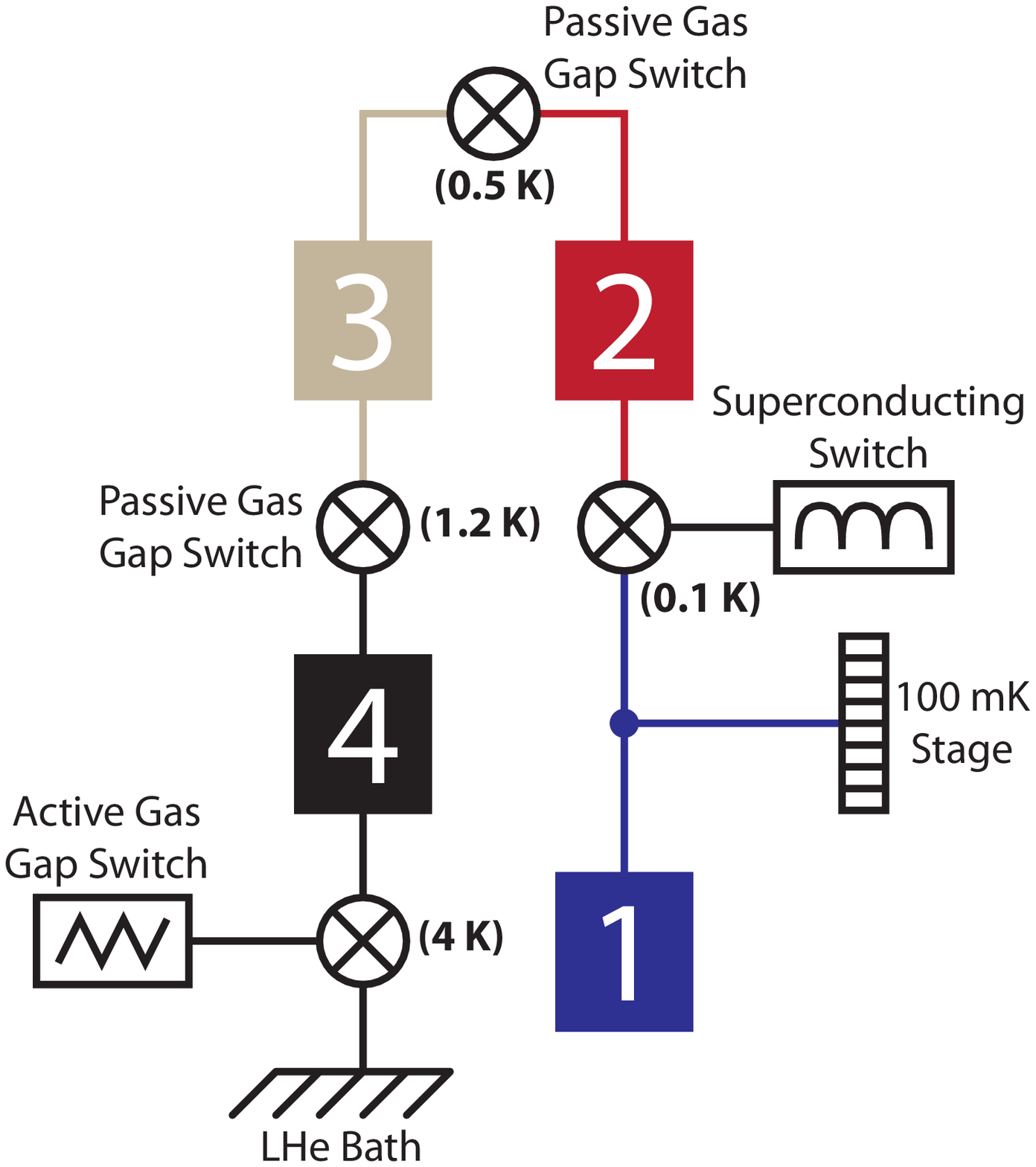}
       \includegraphics[height=7cm]{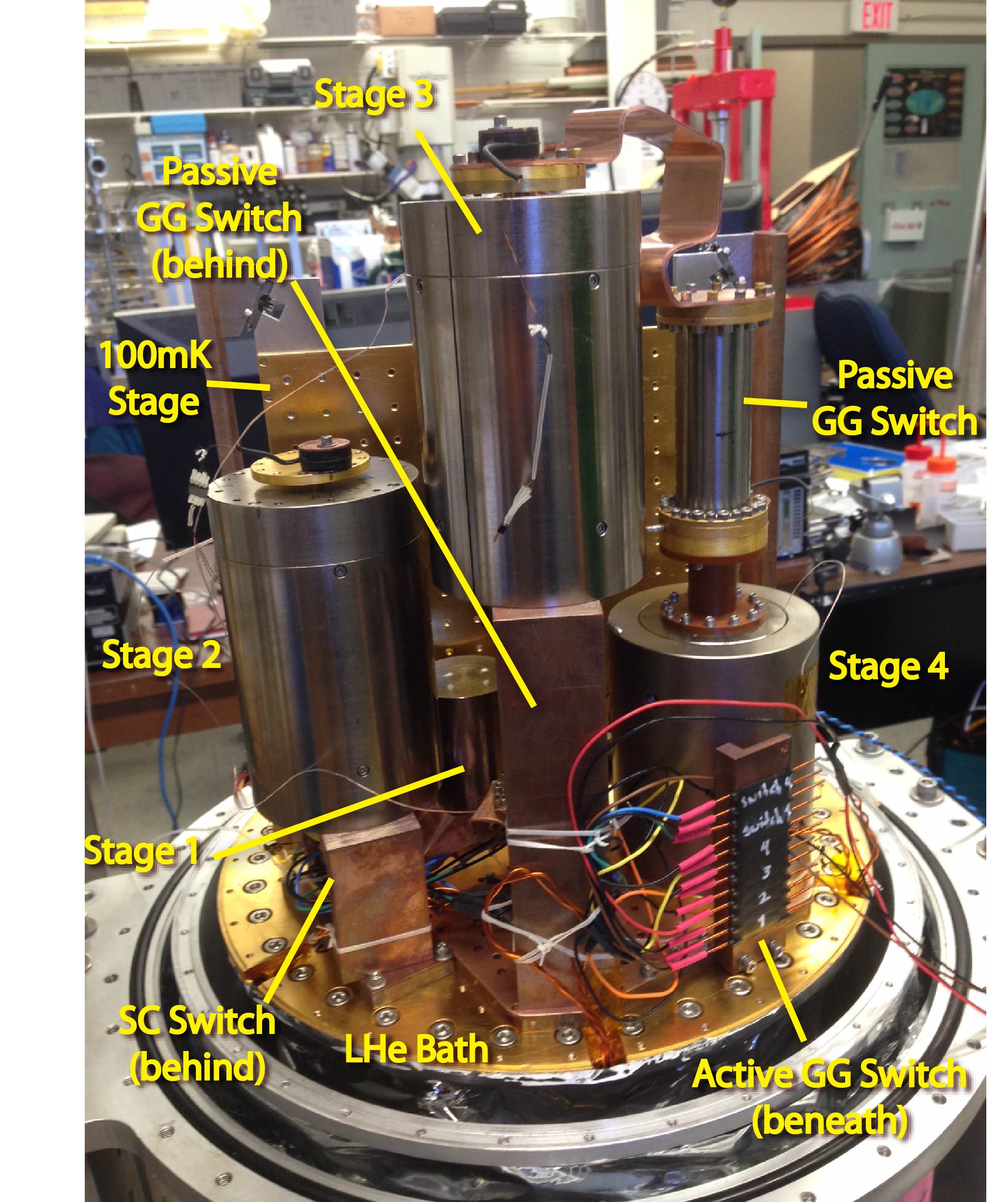}
   \end{center}
   \caption[ADR configuration]{(Left) The 4-stage continuous adiabatic demagnetization refrigerator (ADR) uses 4 salt pills, 4 superconducting solenoids, and 4 switches to operate as a heat pump and provide continuous cooling power to the 100 mK stage. Stage 1 provides a thermal reservoir to stabilize the 100 mK stage. Stages 2 and 3 pump heat from stage 1 to stage 4. Stage 4 provides a virtual 1.5 K pumped LHe bath on the ground. (Right) The flight ADRs have been used extensively in the lab to demonstrate their functionality and to provide a maximally flight-like test bed for detector development.}
   \label{fig:adr}
\end{figure}

The detectors, lenses, analyzer grid, and filters are housed in a vacuum vessel, the ``submarine'', inside the 3500 liter LHe bucket dewar. A continuous-cycle 4-stage adiabatic demagnetization refrigerator\cite{shirron_development_2004} (ADR) cools the detector stage to 100 mK. A series of active and passive heat switches separates the stages of the ADR (Fig.~\ref{fig:adr}). The stages are labeled from 1 (coldest) to 4 (warmest).

Stage 4 couples to the LHe bath through an active gas gap switch and cools the system from the LHe bath temperature to 1.5 K. On the ground at 1 atm, stage 4 cools the system from 4.2 K to 1.5 K and provides a virtual pumped LHe bath; at float, stage 4 is unused, since the LHe bath is already at 1.5 K. Stage 3 couples to stage 4 through a passive gas gap switch and pumps heat from stage 2 to stage 4. Stage 2 is coupled to stage 3 through a second passive gas gap switch and is coupled to stage 1 and the 100 mK stage through an active superconducting switch. Stage 2 provides cooling power to the 100 mK cold stage and stage 1. Stage 1 is coupled directly to the 100 mK stage and serves as a thermal reservoir to hold the 100 mK stage at the desired temperature while the warmer stages are recycling. Once the cycle is completed, stage 1 recharges while stage 2 holds the 100 mK stage at temperature. By appropriately recycling and decoupling the various stages, the ADR can provide up to $\unit[15]{\mu W}$ of continuous cooling power at 100 mK\cite{shirron_development_2004}.

Kevlar suspensions thermally isolate the 100 mK stage on which the 4 detector packages sit. All other components in the submarine will be held at the 1.5 K LHe bath temperature. The SQUID series arrays, ADR assembly, and optical components are thermally coupled to a 1" copper thermal bus bar embedded in a hermetic flange that passes through the submarine shell. The bus bar extends out and down from the submarine to sit in the superfluid bath, providing a high-conductivity thermal pathway to the LHe bath for the components inside the submarine. The flight housekeeping and detector array cables pass through a thermal stop coupled to the thermal bus bar before traveling up a bellows toward the dewar aperture. At the top, a cluster flange splits the tube into many KF ports for routing cables to the appropriate electronics boxes and for evacuating the volume. The MCE cables go to boxes mounted directly onto the four MCEs by MPFPI\footnote{\url{http://mpfpi.com}} hermetic MDMW100 feedthroughs.

The four MCEs require a total of 20 bundles of 50 twisted pairs of 0.005" manganin, each 4 m long, to the submarine. The heat leak through the wires connecting the series arrays to the detectors is minimized, since this heat leak must be absorbed by the ADRs, by using 30 cm of twisted pair 0.003" NbTi wiring with CuNi cladding.

\section{ELECTRONICS}
\label{sec:electronics}


The detector arrays are read out by four sets of Multi-channel Electronics (MCEs) provided by the University of British Columbia. Each MCE is driven by a single external clock provided by the SyncBox so that the measurements are synchronous. For \piper, the science signal is constructed by combining the measurements from the detectors with pointing data and VPM position. It is advantageous if these disparate signals are all sampled synchronously.

The \piper\ electronics are a set of custom boards that are ideally suited to high-precision balloon-borne telescopes. They are low-power compact boards that are synchronized with the MCEs and designed to be run off of DC power supplies such as batteries. A single 3U rack can hold 20 cards in a backplane, each card requiring only $\unit[\sim\!\!1]{W}$ of power. Since all of the cards are housed in the same backplane, a single set of power lines may be used for all of the cards in the backplane. The cards are used to measure all sensors, excluding detectors, in the \piper\ instrument, including the pointing sensors and VPM position sensors, synchronously to within 10 ns.

All of the \piper\ boards are designed in-house around Microchip dsPIC30F5011 programmable microcontrollers\footnote{\url{http://www.microchip.com/wwwproducts/Devices.aspx?product=dsPIC30F5011}}. The dsPIC30F5011 requires an external clock. For lab use, a single crystal oscillator on the master card (see below) is used for all of the boards; for flight use, the SyncBox clock is piped in and the on-board crystal oscillator is de-powered. In either scenario, the same clock is used to drive the dsPIC30F5011 on every board, so the boards cannot drift. On powerup, the boards synchronize to the MCE frame count and begin outputting data. In this way, if there is a power failure, the boards automatically re-synchronize and resume outputting data with no user or flight computer interaction required. \piper\ will use 5 varieties of custom boards: a master card, a fast ADC card, a 4-wire measurement card, an analog in card, and an analog out card.

All of the \piper\ cards are coordinated by a master card. The master card takes in the clock and data lines from the SyncBox, parses the MCE frame count, sends out a synchronization signal to the other cards, and distributes the clock to the other cards. It also communicates on behalf of most cards in the backplane with the outside world via fiber optics, so a single communications channel can control many boards.

A fast ADC card is used to monitor the rapidly changing sensors on the payload, e.g.\ the VPM position sensors and pointing gyros. It is a 32-channel ADC board with a $\unit[~\!\!2880/N_{ch}]{Hz}$ sampling rate\footnote{Exact rate depends on clock settings, but this value is typical.\label{fn:clock}}, where $1 \leq N_{ch} \leq 32$ is the number of channels being monitored. The board has a $\unit[40]{\nicefrac{\mu V}{\sqrt{Hz}}}$ spectral noise density with a $1/f$ knee at 1 mHz.

The 4-wire measurement card is a 16-channel adjustable-gain 4-wire AC resistance bridge. It drives a 16 Hz square wave adjustable current through a load to measure its resistance. Measurements are made at 1 Hz\footref{fn:clock}. The board may be tuned to operate ideally for $\unit[\sim\!\!1]{\Omega}$ loads, $\unit[\sim\!\!10]{k\Omega}$ loads, or diodes. Systematic errors may be calibrated out using the built-in calibration resistors. The card runs off the SyncBox clock and does not generate significant EMI. The board has a fractional resistance sensitivity of $\frac{\Delta R}{R} \simeq \unit[2.5 \times 10^{-5}]{\sqrt{s}}$ for a $\unit[20]{k\Omega}$ resistor biased at 32 nA, typical for a RuOx thermometer.

The PID control card is used to control an ADR. It is paired with a current amplifier to drive the large superconducting solenoid that actuates an ADR. It has functionality built-in to hold at constant temperature or constant current, as well as a linear ramp in temperature or current. A thermometer measured with the same resistance bridge circuitry as the 4-wire card is used as feedback on the ADR temperature. We have demonstrated temperature stability using a PID control card to $<\!\unit[10]{\mu K}$.

The analog in and analog out cards are used for controlling or actuating room temperature thermometers, pressure sensors, relays, and other non-critical sensors and actuators. Each has 32 channels and samples/updates at 1 Hz\footref{fn:clock}.

Communications with the \piper\ electronics are performed entirely over fiber optics. This isolates the sensitive signal channels from digital noise generated by the computers. Additionally, all inputs or outputs to the cards are buffered, further protecting the signal lines. This greatly simplifies the task of preventing noise from coupling into the signal lines, as the only sources are the power lines and the signal line themselves; these noise sources may be mitigated by proper grounding, proper shielding, and filtering.

A primary concern with using a SQUID detector readout is that it is very sensitive to noise coupling into the readout circuit. The \piper\ electronics have been demonstrated to be compatible with an MCE-like SQUID readout and do not noticeably couple into the detector signal or reduce stability.

\section{Conclusion}
\label{sec:conclusion}

\piper\ will achieve the large sky coverage, fast mapping speed, high sensitivity, and good systematic control required to map the CMB B-mode spectrum for all $\ell < 300$ down to $r < 0.007$. This will allow \piper\ to unambiguously test the \bicep2\ results and extend them to explore the physics of inflation. \piper\ will launch the first of its 8 conventional balloon flights in the Fall of 2015 out of Ft. Sumner, NM.




\bibliography{SPIE2014-Montreal}   

\begin{thebibliography}{10}

\bibitem{hu_cmb_1997}
Hu, W. and White, M., ``A {CMB} polarization primer,'' {\em New Astronomy}~{\bf
  2},  323--344 (Oct. 1997).

\bibitem{kamionkowski_statistics_1997}
Kamionkowski, M., Kosowsky, A., and Stebbins, A., ``Statistics of cosmic
  microwave background polarization,'' {\em Phys. Rev. D}~{\bf 55},  7368--7388
  (June 1997).

\bibitem{collaboration_bicep2_2014}
Ade, P. A.~R., Aikin, R.~W., Barkats, D., et~al., ``{BICEP2} i: Detection of
  b-mode polarization at degree angular scales,'' {\em {ArXiv} e-prints}  (Mar.
  2014).

\bibitem{collaboration_planck_2013}
Ade, P. A.~R., Aghanim, N., Armitage-Caplan, C., Arnaud, M., Ashdown, M.,
  Atrio-Barandela, F., Aumont, J., Baccigalupi, C., Banday, A.~J., et~al.,
  ``Planck 2013 results. {XVI.} cosmological parameters,'' {\em {ArXiv}
  e-prints}  (Mar. 2013).

\bibitem{filippini_spider:_2010}
Filippini, J.~P., Ade, P. A.~R., Amiri, M., Benton, S.~J., et~al., ``{SPIDER:}
  a balloon-borne {CMB} polarimeter for large angular scales,'' in [{\em
  Society of Photo-Optical Instrumentation Engineers ({SPIE)} Conference
  Series}{\nolinebreak\hspace{0.1em}]},  {\em Society of Photo-Optical
  Instrumentation Engineers ({SPIE)} Conference Series} {\bf 7741} (July 2010).

\bibitem{reichborn-kjennerud_ebex:_2010}
Reichborn-Kjennerud, B., Aboobaker, A.~M., Ade, P., Aubin, F., Baccigalupi, C.,
  Bao, C., Borrill, J., Cantalupo, C., Chapman, D., Didier, J., Dobbs, M.,
  Grain, J., Grainger, W., Hanany, S., Hillbrand, S., Hubmayr, J., Jaffe, A.,
  Johnson, B., Jones, T., Kisner, T., Klein, J., Korotkov, A., Leach, S., Lee,
  A., Levinson, L., Limon, M., {MacDermid}, K., Matsumura, T., Meng, X.,
  Miller, A., Milligan, M., Pascale, E., Polsgrove, D., Ponthieu, N., Raach,
  K., Sagiv, I., Smecher, G., Stivoli, F., Stompor, R., Tran, H., Tristram, M.,
  Tucker, G.~S., Vinokurov, Y., Yadav, A., Zaldarriaga, M., and Zilic, K.,
  ``{EBEX:} a balloon-borne {CMB} polarization experiment,'' in [{\em Society
  of Photo-Optical Instrumentation Engineers ({SPIE)} Conference
  Series}{\nolinebreak\hspace{0.1em}]},  {\em Society of Photo-Optical
  Instrumentation Engineers ({SPIE)} Conference Series} {\bf 7741} (July 2010).

\bibitem{chuss_variable-delay_2014}
Chuss, D.~T., Eimer, J.~R., Fixsen, D.~J., Hinderks, J., Kogut, A.~J., Lazear,
  J., Mirel, P., Switzer, E., Voellmer, G.~M., and Wollack, E.~J.,
  ``Variable-delay polarization modulators for cryogenic millimeter-wave
  applications,'' {\em {ArXiv} e-prints}  (Mar. 2014).

\bibitem{mortonson_joint_2014}
Mortonson, M.~J. and Seljak, U., ``A joint analysis of planck and {BICEP2} b
  modes including dust polarization uncertainty,'' {\em {ArXiv} e-prints}  (May
  2014).

\bibitem{flauger_toward_2014}
Flauger, R., Hill, J.~C., and Spergel, D.~N., ``Toward an understanding of
  foreground emission in the {BICEP2} region,'' {\em {ArXiv} e-prints}  (May
  2014).

\bibitem{collaboration_planck_2014}
Ade, P. A.~R., Aghanim, N., Alina, D., Alves, M. I.~R., Armitage-Caplan, C.,
  Arnaud, M., Arzoumanian, D., Ashdown, M., Atrio-Barandela, F., et~al.,
  ``Planck intermediate results. {XIX.} an overview of the polarized thermal
  emission from galactic dust,'' {\em {ArXiv} e-prints}  (May 2014).

\bibitem{finkbeiner_extrapolation_1999}
Finkbeiner, D.~P., Davis, M., and Schlegel, D.~J., ``Extrapolation of galactic
  dust emission at 100 microns to cosmic microwave background radiation
  frequencies using {FIRAS},'' {\em {ApJ}}~{\bf 524},  867 (Oct. 1999).

\bibitem{ade_review_2006}
Ade, P. A.~R., Pisano, G., Tucker, C., and Weaver, S., ``A review of metal mesh
  filters,'' in [{\em Society of Photo-Optical Instrumentation Engineers
  ({SPIE)} Conference Series}{\nolinebreak\hspace{0.1em}]},  {\em Society of
  Photo-Optical Instrumentation Engineers ({SPIE)} Conference Series} {\bf
  6275} (July 2006).

\bibitem{eimer_primordial_2010}
Eimer, J.~R., Ade, P. A.~R., Benford, D.~J., Bennett, C.~L., Chuss, D.~T.,
  Fixsen, D.~J., Kogut, A.~J., Mirel, P., Tucker, C.~E., Voellmer, G.~M., and
  Wollack, E.~J., ``The primordial inflation polarization explorer ({PIPER):}
  optical design,'' in [{\em Society of Photo-Optical Instrumentation Engineers
  ({SPIE)} Conference Series}{\nolinebreak\hspace{0.1em}]},  {\em Society of
  Photo-Optical Instrumentation Engineers ({SPIE)} Conference Series} {\bf
  7733} (July 2010).

\bibitem{datta_large-aperture_2013}
Datta, R., Munson, C.~D., Niemack, M.~D., {McMahon}, J.~J., Britton, J.,
  Wollack, E.~J., Beall, J., Devlin, M.~J., Fowler, J., Gallardo, P., Hubmayr,
  J., Irwin, K., Newburgh, L., Nibarger, J.~P., Page, L., Quijada, M.~A.,
  Schmitt, B.~L., Staggs, S.~T., Thornton, R., and Zhang, L., ``Large-aperture
  wide-bandwidth antireflection-coated silicon lenses for millimeter
  wavelengths,'' {\em Appl. Opt.}~{\bf 52},  8747--8758 (Dec. 2013).

\bibitem{chuss_interferometric_2006}
Chuss, D.~T., Wollack, E.~J., Moseley, S.~H., and Novak, G., ``Interferometric
  polarization control,'' {\em Applied Optics}~{\bf 45},  5107--5117 (July
  2006).

\bibitem{chuss_primordial_2010}
Chuss, D.~T., Ade, P. A.~R., Benford, D.~J., Bennett, C.~L., Dotson, J.~L.,
  Eimer, J.~R., Fixsen, D.~J., Halpern, M., Hilton, G., Hinderks, J., Hinshaw,
  G., Irwin, K., Jackson, M.~L., Jah, M.~A., Jethava, N., Jhabvala, C., Kogut,
  A.~J., Lowe, L., {McCullagh}, N., Miller, T., Mirel, P., Moseley, S.~H.,
  Rodriguez, S., Rostem, K., Sharp, E., Staguhn, J.~G., Tucker, C.~E.,
  Voellmer, G.~M., Wollack, E.~J., and Zeng, L., ``The primordial inflation
  polarization explorer ({PIPER)},'' in [{\em Society of Photo-Optical
  Instrumentation Engineers ({SPIE)} Conference
  Series}{\nolinebreak\hspace{0.1em}]},  {\em Society of Photo-Optical
  Instrumentation Engineers ({SPIE)} Conference Series} {\bf 7741} (July 2010).

\bibitem{mainini_improved_2013}
Mainini, R., Minelli, D., Gervasi, M., Boella, G., Sironi, G., Baú, A., Banfi,
  S., Passerini, A., Lucia, A.~D., and Cavaliere, F., ``An improved upper limit
  to the {CMB} circular polarization at large angular scales,'' {\em J. Cosmol.
  Astropart. Phys.}~{\bf 2013},  033 (Aug. 2013).

\bibitem{kusaka_modulation_2014}
Kusaka, A., Essinger-Hileman, T., Appel, J.~W., Gallardo, P., Irwin, K.~D.,
  Jarosik, N., Nolta, M.~R., Page, L.~A., Parker, L.~P., Raghunathan, S.,
  Sievers, J.~L., Simon, S.~M., Staggs, S.~T., and Visnjic, K., ``Modulation of
  {CMB} polarization with a warm rapidly-rotating half-wave plate on the
  atacama b-mode search ({ABS)} instrument,'' {\em Review of Scientific
  Instruments}~{\bf 85},  024501 (Feb. 2014).
\newblock {arXiv:1310.3711} [astro-ph].

\bibitem{kogut_primordial_2012}
Kogut, A., Ade, P. A.~R., Benford, D., Bennett, C.~L., Chuss, D.~T., Dotson,
  J.~L., Eimer, J.~R., Fixsen, D.~J., Halpern, M., Hilton, G., Hinderks, J.,
  Hinshaw, G.~F., Irwin, K., Jhabvala, C., Johnson, B., Lazear, J., Lowe, L.,
  Miller, T., Mirel, P., Moseley, S.~H., Rodriguez, S., Sharp, E., Staguhn,
  J.~G., Tucker, C.~E., Weston, A., and Wollack, E.~J., ``The primordial
  inflation polarization explorer ({PIPER)},'' in [{\em Society of
  Photo-Optical Instrumentation Engineers ({SPIE)} Conference
  Series}{\nolinebreak\hspace{0.1em}]},  {\em Society of Photo-Optical
  Instrumentation Engineers ({SPIE)} Conference Series} {\bf 8452} (Sept.
  2012).

\bibitem{jhabvala_kilopixel_2014}
Jhabvala, C. et~al., ``Kilopixel backshort-under-grid arrays for the primordial
  inflation polarization explorer,'' in [{\em Society of Photo-Optical
  Instrumentation Engineers ({SPIE)} Conference
  Series}{\nolinebreak\hspace{0.1em}]},   {\bf 9153},  127 (June 2014).

\bibitem{chervenak_superconducting_1999}
Chervenak, J.~A., Irwin, K.~D., Grossman, E.~N., Martinis, J.~M., Reintsema,
  C.~D., and Huber, M.~E., ``Superconducting multiplexer for arrays of
  transition edge sensors,'' {\em Applied Physics Letters}~{\bf 74},
  4043--4045 (June 1999).

\bibitem{irwin_-focal-plane_2004}
Irwin, K.~D., Audley, M.~D., Beall, J.~A., Beyer, J., Deiker, S., Doriese, W.,
  Duncan, W., Hilton, G.~C., Holland, W., Reintsema, C.~D., Ullom, J.~N., Vale,
  L.~R., and Xu, Y., ``In-focal-plane {SQUID} multiplexer,'' {\em Nuclear
  Instruments and Methods in Physics Research Section A: Accelerators,
  Spectrometers, Detectors and Associated Equipment}~{\bf 520},  544--547 (Mar.
  2004).

\bibitem{battistelli_functional_2008}
Battistelli, E.~S., Amiri, M., Burger, B., Halpern, M., Knotek, S., Ellis, M.,
  Gao, X., Kelly, D., {MacIntosh}, M., Irwin, K., and Reintsema, C.,
  ``Functional description of read-out electronics for time-domain multiplexed
  bolometers for millimeter and sub-millimeter astronomy,'' {\em J Low Temp
  Phys}~{\bf 151},  908--914 (May 2008).

\bibitem{shirron_development_2004}
Shirron, P., Canavan, E., {DiPirro}, M., Francis, J., Jackson, M., Tuttle, J.,
  King, T., and Grabowski, M., ``Development of a cryogen-free continuous {ADR}
  for the constellation-x mission,'' {\em Cryogenics}~{\bf 44},  581--588 (June
  2004).

\end{thebibliography}
\bibliographystyle{spiebib}   

\end{document}